\documentclass{elsart}
\usepackage{amsfonts, amsmath, amssymb, latexsym}

\journal{Comput.\ Phys.\ Commun.}

\newcommand{\Id}{{\mathrm d}}

\newcommand{\BBN}{{\mathbb{N}}}
\newcommand{\BBZ}{{\mathbb{Z}}}
\newcommand{\Reduce}{\textsc{Reduce}}
\newcommand{\Crack}{\textsc{Crack}}
\newcommand{\SsTools}{\textsc{SsTools}}   
\newcommand{\sym}{\mathop{\rm sym}\nolimits}

\newcommand{\cE}{\mathcal{E}}

\newcommand{\cR}{\mathcal{R}}

\newcommand{\vph}{\varphi}

\newcommand{\sd}{\mathcal{D}}
\newcommand{\oh}{\tfrac{1}{2}}

\DeclareMathOperator{\Lin}{Lin}

\newcommand{\by}[1]{\textrm{{#1}}}
\newcommand{\jour}[1]{\textrm{{#1}}}
\newcommand{\vol}[1]{\textrm{{#1}}}
\newcommand{\book}[1]{\textrm{{#1}}}

\begin{document}
\begin{frontmatter}
\title{Classification of integrable super\/-\/systems
using the SsTools environment\thanksref{CPCSC}}
\thanks[CPCSC]{To appear in: Computer Physics Communications (2007)
Computer programs in physics (submitted October 2, 2006, 
in final form January 19, 2007).%
}
\author{A.~V.~Kiselev\corauthref{corrauth}},
\ead{arthemy@mpim-bonn.mpg.de}
\address{Max Planck Institute for Mathematics,
Vivatsgasse 7, D-53111 Bonn, Germany.}
\corauth[corrauth]{\textit{Address for correspondence}:
Department of Higher Mathematics,\ Ivanovo State
Power University,\ Rabfakovskaya str.~34, Ivanovo, 153003
Russia.}
\author{T.~Wolf}
\ead{twolf@brocku.ca}
\address{Department of Mathematics, Brock University,
500 Glenridge ave.,  St.~Catharines, Ontario, Canada L2S~3A1.}
\begin{abstract}
A classification problem is proposed for supersymmetric
evolutionary PDE that satisfy
the assumptions of nonlinearity, nondegeneracy, and homogeneity.
Four classes of nonlinear coupled boson\/-\/fermion
systems are discovered under the weighting assumption
$|f|=|b|=|D_t|=\oh$.
The syntax of the \Reduce\ package \SsTools, which was used for
intermediate computations, and the applicability of its procedures to
the calculus of super\/-\/PDE are described.
\end{abstract}
\end{frontmatter}

\noindent\textbf{PROGRAM SUMMARY}

\noindent\textit{Title of program}: \SsTools\ %

\noindent\textit{Catalogue number}: \texttt{xxxx}

\noindent\textit{Program obtainable from}: CPC~Program Library, Queen's
University of Belfast, N.~Ireland;
see also~[1]

\noindent\textit{Licensing provisions}: none

\noindent\textit{Computer for which the program is designed and others
on which it has been tested}:\\
\textit{Computers}: (i) IBM~PC, (ii) cluster\\
\textit{Installations}: (i) Brock University, St.\ Catharines, Ontario,
Canada L2S~3A1; (ii) SHARCNET \texttt{http://www.sharcnet.ca}

\noindent\textit{Operating system under which the program has been
tested}: LINUX

\noindent\textit{Programming language used}: REDUCE~3.7, REDUCE~3.8

\noindent\textit{Memory required to execute with typical data}: problem
dependent (10\,Mb -- 1Gb), typical working size $<$ 100~Mb

\noindent\textit{No.\ of bits in a word}: $32, 64$

\noindent\textit{No.\ of processors used}: (i) $1$; (ii) multiple

\noindent\textit{Peripherals used}: no

\noindent\textit{Has the code been vectorized}? no

\noindent\textit{No.\ of lines in distributed program including
documentation file, test data, etc.}: 2485

\noindent\textit{Distribution format}: ASCII

\noindent\textit{Online access and tutorials}: \texttt{http://lie.math.brocku.ca/crack/susy/}


\noindent\textit{Nature of physical problem}:
The program allows the classification
of $N\geq1$ supersymmetric nonlinear scaling\/-\/invariant
evolution equations $\{f_t=\varphi^f$, $b_t=\varphi^b\}$
that admit infinitely many local symmetries propagated by recursion
operators; here $b(x,t;\theta)$ is the set of
bosonic super\/-\/fields and $f(x,t;\theta)$ are fermionic
super\/-\/fields.

\noindent\textit{Method of solution}:
First, (half-)\/integer weights $|f|$, $|b|$, $\ldots$,
$|D_t|$, $|D_x|\equiv1$ are assigned
to all variables and derivatives and then
pairs of commuting flows that are homogeneous w.r.t.\ these
weights are constructed.
Secondly, the seeds of higher symmetry sequences [2] for the systems
are sorted out,
and finally the recursion operators that generate
the symmetries are obtained~[3].
The intermediate algebraic systems upon the undetermined coefficients
are solved by using~[4].

\noindent\textit{Restrictions on the complexity of the problem}:
Computation of symmetries of high differential order for very large
evolutionary systems may cause memory restrictions. Additional
si\-ze/ti\-me restrictions may occur if the homogeneity weights of some
super\/-\/fields are non\/-\/positive, see section~\ref{pCalcRec} of the
Long Write\/-\/Up.

\noindent\textit{Typical running time}:
depends on the size and complexity of the input system
and varies between seconds and minutes.

\noindent\textit{Unusual features of the program}:
\SsTools\ has been extensively tested using hundreds of PDE systems
within three years on UNIX\/-\/based PC\/-\/machines.
\SsTools\ is applicable to the computation of symmetries,
conservation laws, and Hamiltonian structures for
$N\geq1$ evolutionary super\/-\/systems with any~$N$.
\SsTools\ is also useful for performing extensive arithmetic 
of general nature including differentiations of super\/-\/field expressions.

\noindent\textit{References}:\\
{\small
{[1]} 
\texttt{http://lie.math.brocku.ca/crack/susy/sstools.red};
The support package \Crack\ is obtained from
\texttt{http://lie.math.brocku.ca/crack/src/crack.tar.gz}\\
{[2]} 
\by{P.J.~Olver}, \book{Applications of Lie groups to differential
equations}, $2$nd ed.,  
Springer, Berlin, 1993.\\
{[3]} 
\by{I. S. Krasil'shchik, P. H. M. Kersten},
\book{Symmetries and recursion operators for classical and
supersymmetric differential equations}, Kluwer, 
Dordrecht, 
2000.\\
{[4]} 
\by{T. Wolf}, Applications of \textsc{Crack} in the classification of
integrable systems, \book{CRM Proc.\ Lecture Notes} \vol{37}
(2004), 283--300. 

}

\noindent\textbf{LONG WRITE\/--\/UP}
\subsection*{Introduction}
The principle of symmetry belongs to the foundations of modern
mathematical physics. The differential equations that constitute
\emph{integrable}~\cite{MiShSok} models practically always admit symmetry
transformations and, reciprocally, new classes of integrable phenomena
in physics are obtained by postulating some symmetry invariance.
The presence of symmetry transformations in a system
yields two types of explicit solutions: those which are invariant under
a transformation (sub)\/group and, secondly, the
solutions obtained by propagating a known solution by the same group. The
two schemes for constructing new solutions of PDE are crucial for
systems of super\/-\/equations of mathematical physics (e.g.,
supergravity models); these equations involve commuting
(bosonic, or `even') and anticommuting (fermionic, or `odd')
independent variables and/or unknown functions. Indeed,
other methods for solving nonlinear equations need a special adaptation
to the super\/-\/field setting, see e.g.~\cite{Carstea};
another approach to integrability of supersymmetric equations has been
investigated in~\cite{KuperSS}.
Hence the symmetry considerations~\cite{MiShSok}, which are based on the
computation of infinitesimal symmetry generators for PDE, become highly
important. The arising computational problems are unmanageable without
computer algebra that permits handling relevant systems.

In the literature it has been observed
(see~\cite{MiShSok,Tsuchida} and references therein)
that the principal phenomena in nature
are governed by systems of PDE that admit higher symmetries, that is,
the symmetry transformations that involve higher order derivatives of
the unknown functions. Simultaneously, when dealing with supersymmetric
models of theoretical physics, it is often hard to predict whether a
certain mathematical approximation will be truly integrable or not.
Therefore we apply the symbolic computational approach to the physical
problem of classifying the systems that exhibit necessary integrability
features. It must be noted that the very idea to filter out integrable
cases using the presence of `many' symmetries is widely accepted in the
computer branch of modern mathematical physics,
see e.g.\ \cite{Tsuchida,JKKersten}. 
These systems are called symmetry integrable~\cite{MiShSok,Tsuchida});
in some cases, they can be transformed to exactly solvable equations or
their extensions. In view of this classification task, we analyzed
fermionic extensions of the Burgers and Boussinesq equations and
related the former with evolutionary systems on associative algebras
in~\cite{Kiev2005}.

This paper is organized as follows.
First we formulate the axioms of the classification problem for $N=1$
supersymmetric systems of evolutionary~PDE.
In sections~\ref{SecCommFlows} and~\ref{pCalcRec} we describe the two
modes of the procedure \verb+ssym+ in the package \SsTools\
that allow, respectively, finding unordered pairs of commuting flows
and the computation of symmetries for previously found systems.
Also, in section~\ref{pCalcRec} we review a geometric (coordinate\/-\/free)
algorithm for constructing recursion operators.
However, we refer to~\cite{JKKersten,Olver} for basic notions and concepts
in the geometry of (super)\/PDE, see
also~\cite{Wintern,FPM,Lstar,WolfCrack} and references therein.
The principal result of this paper is that there exist only
four nonlinear coupled boson\/-\/fermion
systems that satisfy the axioms and
the weighting $|f|=|b|=|D_t|=\tfrac{1}{2}$.
Section~\ref{SecClass} concludes with
the classification of their recursion operators.
In the following sections we investigate properties of these four
systems, giving simultaneously the syntax and describing applicability
of \SsTools\ subroutines developed for the calculation of
the scaling weights,
symmetries, linearizations, conservation laws needed for introducing
the nonlocalities, and recursion operators. Finally, 
sample runs of \SsTools\ are given.

\section{The classification problem}\label{SecClass}
Let us introduce some notation.
We denote by $\theta$ the super\/-\/variable
and we put $\sd\equiv D_\theta+\theta\,D_x$
such that $\sd^2=D_x$ and $[\sd{},\sd{}]=2D_x$;
here $D_\theta$ and $D_x$ are the 
derivatives w.r.t.\ $\theta$ and
$x$, respectively. Throughout this text, the operator~$\sd{}$ acts on
the succeeding super\/-\/field unless stated otherwise explicitly.

The package \SsTools\ in \Reduce\ (ver.~3.7 or higher)
is designed for calculus on super\/-\/PDE. 
It was implemented for solving the classification problem for symmetry
integrable $N\geq1$ super\/-\/PDE and fermionic extensions of the
purely bosonic evolution equations.
Compared to other programs listed in the overview \cite{WHER}, it is able to work with (super)\/PDE
and compared to programs mentioned in \cite{Hussin,Kersten,Popowicz} the package
\SsTools\ has not only the ability to compute higher symmetries for a given
system but also to solve the non\/-\/linear problem of computing polynomial systems
and their higher symmetries \emph{at once} if certain
homogeneity weights are specified;
here we also note the program~\cite{Krivonos} for computations
performed in presence of the supersymmetry.
In \SsTools, the following list of seven axioms
was postulated:
\begin{enumerate}\label{pAxioms}
\item
each system of equations $\binom{f_t}{b_t}$ admits at least one higher
symmetry $\binom{f_s}{b_s}$;
\item
all equations are spatial
translation invariant and do not depend on the
time $t$ explicitly;
\item\label{ax3}
none of the evolution equations involves only one field
and hence none of the right\/-\/hand sides vanishes;
\item\label{ax4}
at least one of the right\/-\/hand sides in either the
evolution equation or its symmetry is nonlinear;
\item\label{AxN0N1}
at least one equation in a system or at least one component of its
symmetry contains a fermionic field or
the super\/-\/derivative~$\sd{}$;
\item\label{AxScale}
the equations are scaling invariant: their right\/-\/hand sides are
differential polynomials homogeneous
with respect to a set of (half-)\/integer
weights\footnote{The \emph{weights} $|\cdot|$ take all variables
(e.g., a fermionic super\/-\/field $f$) and derivations 
(e.g., $D_t\equiv \mathrm{d}/\mathrm{d}t$) to (half-)\/integers
($|f|$ and $|D_t|\in\tfrac{1}{2}\BBZ$, respectively) such that
the weight of any product is the sum of weights of the factors
(hence $|f_t|=|D_t(f)|=|D_t|+|f|$); 
only homogeneous algebraic expressions with equal weights of the
components are considered from now on.}
$|\sd{}|\equiv\oh$, $|D_x|\equiv1$,
$|D_t|>0$, $|f|$, $|b|>0$;
we also assume that the positive weight
of $D_s$ that occurs through $\binom{f_s}{b_s}$ is (half\/-)\/integer.
\end{enumerate}
The time $t$ and the parameter $s$ along the integral
trajectories of the symmetry fields
can either assign the bosonic flows $b_t$, $b_s$ to $b$ and the
fermionic flows $f_t$, $f_s$ to $f$ or reverse the parities, that is,
$b_{\bar t}$ or $b_{\bar s}$ become fermionic and $f_{\bar t}$ or
$f_{\bar s}$ are bosonic.
Here we use the notation $\bar{t}$ or~$\bar{s}$ if
the parities of $f_{\bar{t}}$, $b_{\bar{t}}$ or
$f_{\bar{s}}$, $b_{\bar{s}}$
are opposite to the parities of $f$ and~$b$, respectively.

\subsection{Computation of commuting flows}\label{SsTools1}\label{SecCommFlows}
The procedure \verb+ssym+ for computing symmetries, which is the
central procedure\footnote{The command \texttt{sshelp();} provides a
detailed user's guide on the package including examples that illustrate
admissible combinations of the flags in the procedure calls. All
package subroutines are described in this paper where appropriate.
Online access to \SsTools\ without any need to install the computer algebra
system is provided through \texttt{http://lie.math.brocku.ca/crack/susy}.}
of the \SsTools\ package, can be used in two modes.
The first mode of \verb+ssym+ developed together with
W.\,Neun was used for finding
$N=1$ supersymmetric
coupled boson\/-\/fermion
evolutionary systems that satisfy the above axioms.
The axioms are incorporated
in the ansatz generators. Hence, having the
weights $|f|$, $|b|$, $|t|$ or~$|\bar{t}|$, and $|s|$ or~$|\bar{s}|$
specified, we face the problem of constructing a pair of commuting
flows $(f,b)_{t(\text{or }\bar{t})}$ and
$(f,b)_{s(\text{or }\bar{s})}$. The procedure
\texttt{ssym} generates the homogeneous ansatz with undetermined
coefficients in both flows, and calls the program \Crack\ \cite{WolfCrack}
for solving automatically the overdetermined algebraic systems.
The standard call in the first mode is
\begin{verbatim}
 ssym(N,tw,sw,afwlist,abwlist,eqnlist,fl,inelist,flags);
\end{verbatim}
where\\
\verb" N       "$\ldots$ the number of superfields $\theta^i$;\\
\verb" tw      "$\ldots$ $2$ times the weight of $D_t$;\\
\verb" sw      "$\ldots$ $2$ times the weight of $D_s$;\\
\verb" afwlist "$\ldots$ list of weights of the fermion fields
                \verb"f(1)", \verb"f(2)",$\ldots$, \verb"f(nf)";\\
\verb" abwlist "$\ldots$ list of weights of the boson fields
            \verb"b(1)", \verb"b(2)",$\ldots$, \verb"b(nb)";\\
\verb" eqnlist "$\ldots$ list of extra conditions on the undetermined coefficients;\\
\verb" fl      "$\ldots$ extra unknowns in \verb"eqnlist" to be determined;\\
\verb" inelist "$\ldots$ \parbox[t]{11.2cm}{a list, each element of it is a non-zero expression or
                a list with at least one of its elements being non-zero;}\\[1pt]
\verb" flags   "$\ldots$ \parbox[t]{12.7cm}{\texttt{init}: only initialization of global data,\\
            \texttt{zerocoeff}: all coefficients ${}=0$ which do not appear in \texttt{inelist},\\
            \texttt{tpar}: if the time variable $t$ changes parity,\\
            \texttt{spar}: if the symmetry variable $s$ changes parity.}

\smallskip\noindent%
The nonnegative integer~$N$ is the number of super\/-\/fields~$\theta^i$ such that $\mathcal{D}_i\equiv D_{\theta^i}+\theta^i\,D_x$ are the respective super\/-\/derivatives, $\mathcal{D}_i^2=D_x$. The computer notation for $\mathcal{D}_i$ is \verb"d(i,"$
\ldots$\verb")", thus opening the opportunity to work with $N\geq1$ under \SsTools\ doing the calculus of super\/-\/fields;
here we also note that the syntax of $D_x$ is \verb+df(..,x)+ and similarly for~$D_t$.
In addition, \SsTools\ permits the investigation of 
non\/-\/supersymmetric systems (with~$N=0$, which is not
needed for the four systems within the classification below).
What may be of interest to the specialists working with super\/-\/calculus and super\/-\/PDE is that after loading \SsTools\ the simplification of polynomial expressions which
involve the anti\/-\/commuting derivations~$\mathcal{D}_i$ and fields~$f^i$ is performed automatically, taking care of all the minus signs that arise in multiplications and derivatives.

The numbers \verb"nf" of fermion fields and \verb"nb" of boson fields
are determined automatically through the number of elements in the
lists \verb"afwlist" and \verb"abwlist".
The input list \verb"eqnlist" can contain substitutions, like
\verb"p2=1", or expressions which are to be set to zero, like
\verb"p3*r4+p2*r3". Conditions as \verb"p2=1"
are executed instantly at the time of formulating the ansatz
for the pair of commuting flows, while
expressions without \verb"=" sign are added as conditions  $\ldots$\verb"=0"
to the other equations when calling \Crack.
Typically, one would run the program first with the flag \texttt{init}
to see which ansatz for the system and its symmetry is generated and then
start \verb"ssym" again without \texttt{init} but with the option
to add extra conditions on the unknown coefficients either through the
flag \texttt{zerocoeff} or through specific extra conditions
in \verb"eqnlist" or entries in \verb"inelist".

Using the bounds $0<|f|$, $|b|\leq5$ and $0<|D_t|<|D_s|\leq 5$,
we constructed the experimental database~\cite{SUSY}, which
contains $1830$ equations such that $N\leq2$
(the duplicate PDEs that appeared owing to possible
non\/-\/uniqueness of the weights are not counted) and their $4153$
symmetries (plus the translations along $x$ and $t$, and plus the
scalings whose number is in fact infinite).
In this paper we investigate properties of the
systems for which the weights of the
times $t$, $\bar{t}$ are $|t|=|\bar{t}|=-\oh\Leftrightarrow|D_t|=|D_{\bar{t}}|=\oh$, assuming further that
$|f|=|b|=\oh$ (the weights may not be uniquely defined).
Having fixed the weights and using \SsTools,
we discovered four systems that satisfy the set of axioms;
they 
are classified as follows, see Table~\ref{Tab1} on p.~\pageref{Tab1} below.
System~\eqref{DoubleLayer} is related to the fermionic extension of the Burgers equation~\cite{Kiev2005}. Also, we get a super\/-\/field
representation~\eqref{BurgSystem} of the Burgers
equation~\eqref{BurgersInverse} and discuss its properties which are revealed only by the presence of super\/-\/variables.
Further, there is a multiplet of $14$ systems~\eqref{Quad} with the
parity preserving time~$t$; these systems possess
$s$\/-\/symmetries and higher $\bar{s}$\/-\/supersymmetry flows.
Finally, we obtain a
unique system~\eqref{stPar} which has the parity reversing times
$\bar{t}$ and~$\bar{s}$; using this system, we illustrate
the process of calculating the recursion operators in the
test output. 

\subsection{Reconstruction of recursion operators}\label{pCalcRec}
The second application of \textsc{SsTools}\ is the construction of
symmetries, conservation laws, and recursion operators for the symmetry algebras of \emph{given} systems, i.e.\ systems~$\cE$ that have been obtained before
by \verb+ssym+ and placed in the database~\cite{SUSY}.
The method of Cartan's forms~\cite{JKKersten} for the recursion
operators is
applied, see~\cite{Kiev2005} for examples based on the
notation introduced below.
Within this approach, the recursions~$\cR$
are regarded as symmetries $F_{s_R}=\cR^f$, $B_{s_R}=\cR^b$ of the
linearized equations~$\Lin\cE$. Namely, we ignore the differential
function structure of the symmetry flows $f_s=F$, $b_s=B$ and
regard $F$ and $B$ as the components of solutions of the
linearized equations~$\Lin\cE$, which are defined in the next
paragraph. The expressions $\cR=R(F,B)$ are recursion operators
for~$\cE$ if each $\cR$ satisfies the linearized equation~$\Lin\cE$
again and if they are linear w.r.t.\ $F$, $B$, and their derivatives.

The understanding of linearized systems~$\Lin\cE$ from a computational
viewpoint is as follows; we consider the differential polynomial case
since this is what we and \SsTools\ deal with. Given a system~$\cE$,
formally assign the new `linearized' fields $F^i={}$\texttt{f(nf+i)}
and $B^j={}$\texttt{b(nb+j)} to $f^i={}$\texttt{f(i)} and
$b^j={}$\texttt{b(j)}, respectively, with $i=1$,$\ldots$,\texttt{nf}
and $j=1$,$\ldots$,\texttt{nb}.
Pass through all equations, and whenever a power of a derivative of a
variable $f^i$ or $b^j$ is met, differentiate (in the usual sense) this
power with respect to its base, multiply the result from the right by
the same order derivative of $F^i$ or $B^j$, respectively, and insert
the product in the position where the power of the derivative was met.
Now proceed by the Leibnitz rule. The final result, when all equations
in the system~$\cE$ are processed, is the linearized system~$\Lin\cE$.
For example, the linearized counterpart of $b_t=b(\sd{f})^2$ is
$B_t=B\cdot(\sd{f})^2+2b\sd{f}\cdot\sd{F}$.

The linearization~$\Lin\cE$ for a system of evolution equations~$\cE$
is obtained using the procedure \verb"linearize":
\begin{verbatim}
 linearize(pdes,nf,nb);
\end{verbatim}
where\\
\verb" pdes "$\ldots$ list of equations with first order derivatives in l.h.s.;\\
\verb" nf   "$\ldots$ number of the fermion fields \verb"f(1)"$,\ldots$,\verb"f(nf)";\\
\verb" nb   "$\ldots$ number of the boson fields \verb"b(1)"$,\ldots$,\verb"b(nb)".

\noindent%
The values \verb"nf" and \verb"nb" may be greater than one
for the coupled boson\/-\/fermion systems with unique $f$ and $b$ (only such systems are studied throughout this paper) because
some additional variables (the nonlocalities,
see~\eqref{EqNewNonlocal} for a detailed 
explanation) can have been added previously by hand using the method of
trivializing conservation laws.
The actual number of fields is doubled by a call of \verb"linearize",
and in the sequel we always denote these new `linearization variables'
by the respective capital letters.

Now we describe the syntax of a call in the second mode of using procedure \verb+ssym+. In this mode symmetries are computed
(which can be interpreted later as recursion
operators) for a given evolutionary system \verb"pdes".
The call is
\begin{verbatim}
 depend {f(1),f(2),..,f(nf),b(1),b(2),..,b(nb)},x,t;
 max_deg:=..;
 hom_wei:={{sw_1,afwlist_1,abwlist_1},..};
 ssym(N,tw,sw,afwlist,abwlist,pdes,fl,inelist,flags);
\end{verbatim}
where this time we use the actual numbers of fields (they may have
increased after \verb"linearize" produced a bigger system
by assigning new $F^i$ to $f^i$ and $B^j$ to $b^j$).
We now have\\
\verb" pdes  "$\ldots$ \parbox[t]{11.2cm}{the list of equations \texttt{df($\ldots$,t)=$\ldots$}
specifying the (linearized)
          system for which symmetries are to be computed, as well as
          substitution rules \texttt{$\ldots$=>$\ldots$} and other conditions \texttt{$\ldots$=$\ldots$}
          (see Remark~\ref{RemSubst} below);}\\[1pt]
\verb" flags "$\ldots$ \parbox[t]{11.2cm}{the list of flags now includes:\\
          \texttt{lin}: the symmetry to be determined
          is linear w.r.t.\ the fields \texttt{f(i)}, \texttt{b(j)}
          such that $i>{}$\texttt{nf}${}/2$ and $j>{}$\texttt{nb}${}/2$;\\
          \texttt{filter}: if present this flag indicates that the
          homogeneities listed in \texttt{hom\_wei} have to be used in addition;\\
          \texttt{tpar}, \texttt{spar}: $D_{\bar{t}}$ and $D_{\bar{s}}$, respectively,
          is parity changing.}

\smallskip\noindent%
If the right\/-\/hand sides in \verb"pdes" involve any
constants which should be computed such
that the higher symmetries do exist, then these constants are
listed in the input list \verb"fl", like
\verb"{p1,p2,"$\ldots$\verb"}" otherwise \verb"fl" is \verb"{}".
This may be needed when we suspect the presence of a multiplet of
systems parameterized by discrete admissible values of these
coefficients~\texttt{p1},\texttt{p2},$\ldots$.
If the constant coefficients are not given in the list
\verb"fl", then they are treated as
independent parameters and only symmetries (and not these constants)
are determined for generic values of the constant coefficients.

Next, if a boson's weight is non-\/positive or a fermion's weight is negative
then the global variable \verb"max_deg" must have a positive
integer value, which is the highest power of such fields and their
derivatives in any ansatz that the program will generate.

\begin{rem}\label{RemSubst}\upshape
The key for procedure \verb+ssym+ to know whether to compute
symmetries for the \emph{given} system (this is mode~$2$) or to {make an ansatz} for the
\emph{yet unknown}
system with a symmetry (hence, for a pair of commuting flows within mode~$1$)
is the first element of the input list \verb+pdes+.  If this is
of type $\ldots$\verb+=+$\ldots$ or of type
$\ldots$\verb+=>+$\ldots$ {and} if the left hand side
is \verb+df(+$\ldots$\verb")" or \verb+d(+$\ldots$\verb+)+, then:
\begin{itemize}
\item
All elements of \verb+pdes+
of the form \verb+df(+$\ldots$\verb+,t)=+$\ldots$ are collected and interpreted as the system for
which symmetries are to be computed.
\item
All other elements of \verb+pdes+ of the form $\ldots$\verb+=>+$\ldots$ are assumed to have
\verb+df(+$\ldots$\verb+)+ or \verb+d(+$\ldots$\verb+)+ on the left\/-\/hand side. These substitution rules
and differential consequences of them are used to
simplify right\/-\/hand sides of the system, in making the ansatz for the symmetry, and in further computations.
\item
All other elements of type $\ldots$\verb+=+$\ldots$ or elements that are simple expressions (which are set to $0$)
are interpreted as extra conditions on the undetermined coefficients.
\end{itemize}


There are two types of relations to be provided in \verb+pdes+ which is the reason why substitutions \verb"=>" are needed
in addition to the usual relations~\verb"=". First, the very
algorithm~\cite{JKKersten,Lstar} for finding recursions of symmetry
algebras of equations~$\cE$ suggests to treat the recursions as
symmetries of the linearized systems~$\Lin\cE$, taking into account
the original equations (that is, replacing their left\/-\/hand sides
with their right\/-\/hand sides) and differential consequences of them.
The linearized system, which involves time derivatives of Cartan's
forms, is included in the list \verb+pdes+ in form of relations
\verb+df(f(i),t)=+$\ldots$, $i>\mathtt{nf}/2$ and
\verb+df(b(j),t)=+$\ldots$, $j>\mathtt{nb}/2$.

Secondly, we recall that the recursion operators are frequently
nonlocal, thus requiring introduction of new nonlocal variables for
the systems~$\cE$, see~\cite{JKKersten,Kiev2005} for many examples.
To this end, the systems are supplemented by differential relations
(see~\eqref{EqNewNonlocal} in
section~\ref{SecEqns}) which specify the new nonlocal
dependent variables. Then these additional relations and their linearized
counterparts are substituted but their symmetry flows are discarded,
which is equivalent to setting them to zero.
These substitution rules are included in \verb+pdes+ using \verb+=>+ in the form
\verb+df(f(i),x)=>+$\ldots$, \verb+d(1,f(i))=>+$\ldots$ and
\verb+df(b(j),x)=>+$\ldots$, \verb+d(1,b(j))=>+$\ldots$.

The order in which the \verb+=+ and \verb+=>+ relations   occur in \verb"pdes" is inessential.
\end{rem}

Suppose the system in \verb"pdes" is homogeneous w.r.t.\ several
sets of homogeneity weights
simultaneously, that is, it admits several scaling symmetries at once.
One then wants to find a symmetry that obeys the same sets of
homogeneity weights for the super\/-\/fields, each set supplemented
by the expected weight of the parameter~$s$. This is done as follows.
An arbitrarily chosen (see a comment below)
set of the weights \verb+tw,sw+ for the times $t,s$ and
\verb+afwlist,abwlist+ for the super\/-\/fields $f_i,b_j$ is specified among the arguments of \verb"ssym",
while the other weights are
given through the global variable \verb"hom_wei". This is
a list of lists \verb"{sw_i,afwlist_i,abwlist_i}"
with the additional sets of homogeneity weights of the parameter~$s$
and the dependent variables, respectively.
The flag \verb+filter+ causes computation only of those
symmetries that satisfy all additional homogeneities encoded in the global variable \verb"hom_wei" as well.

The choice of the primary set of weights placed in the \verb"ssym"
arguments is arbitrary with the following comment: if all bosons'
weights are positive and all fermions' weights are nonnegative in the
primary set, then the bound \verb"max_deg" is not applied even if
\verb"hom_wei" contains zero or negative weights of any variables.
Indeed, in this case the primary weights already guarantee a finite
number of terms in the ansatz for the symmetry, and the extra
homogeneities in \verb"hom_wei" act as an additional filter.  The
global bound \verb"max_deg" is needed if for each set of homogeneity
weights either one boson's weight is non\/-\/positive or one fermion's
weight is negative.

\begin{rem}\upshape
The (non)uniqueness of the weights
for a given system can be checked
within some range
by using the procedure \verb"wgts":
\begin{verbatim}
 wgts(pdes,nf,nb,maxwt);
\end{verbatim}
where\\
\verb" pdes  "$\ldots$ list of equations
       \verb+df(+$\ldots$\verb+,t) = +$\ldots$;\\
\verb" nf    "$\ldots$ number of the fermion fields \verb"f(1)",$\ldots$,\verb"f(nf)";\\
\verb" nb    "$\ldots$ number of the boson fields \verb"b(1)",$\ldots$,\verb"b(nb)";\\
\verb" maxwt "$\ldots$ only weights with a total sum $\leq$ \verb"maxwt" are to be listed.

\noindent%
The weights are scaled such that the weight of
$D_x$ is always $1$ and the weight of $\sd{}$ is $\oh$.
Note that the computer representation of the weights is twice
the notation used throughout the paper to avoid half-integer values.
This convention is usual.
\end{rem}

\begin{rem}\label{Breadth}\upshape
The systems that admit several scaling symmetries and hence are
homogeneous w.r.t.\ different weights allow to apply the breadth
search method for recursions, which is the following.
Let a recursion of weight $|s_R|$ w.r.t.\ a particular set of weights
for the super\/-\/fields $f$, $b$ and the time $t$
be known. Now, recalculate its weight $|s'_R|$ w.r.t.\ another set
of homogeneity weights
$|f|$, $|b|$
and then find all recursion operators of weight~$|s'_R|$. The list of
solutions will incorporate the known recursion and, possibly, other
operators. Generally, their weights will be different from the weight
of the original recursion w.r.t.\ the initial set of weights.
Hence we repeat the reasonings for each new operator and
thus select the weights $|s_R|$ such that the recursions exist.
This method is a serious instrument for checking consistency
of calculations and elimination of errors since the existence of a solution is always guaranteed. We used it while testing the second mode of \verb"ssym" in the \textsc{SsTools} package.
An example of implementing it is given in the 
test output, where
the recursion~\eqref{BreadthRec} for the multiple homogeneous
$\alpha=2$ system~\eqref{Quad} is constructed.
\end{rem}

\subsection{The classification of systems and recursions}\label{SecEqns}
Let us introduce some notation which will simplify classification
of recursion operators for the systems at hand.
Assume $\cR$ is a recursion for an equation and consider the symbol
${}_{\text{ord}}^{\text{layers}}\cR_{\text{weight}}^{\sharp}$.
The subscripts `ord' and `weight' denote the differential order and the
weight of the recursion $\cR$, respectively.

Recall further that the nonlocal variables are defined for $N=1$
super\/-\/equ\-a\-ti\-ons $\cE$ by
trivializing~\cite{JKKersten,Kiev2005,FPM} conserved currents, which
are of the form $D_t(\rho)+\sd{Q}\doteq0$; here the right\/-\/hand side
vanishes by virtue
($\doteq$) of the system~$\cE$ and all possible differential consequences from it.
The standard procedure~\cite{JKKersten} suggests that this relation is
the compatibility condition for a new `nonlocal' variable (the
nonlocality), say $w$, whose derivatives are set to
\begin{equation}\label{EqNewNonlocal}
\sd{w}\mathrel{{:}{=}}\rho \text{ and } w_t\mathrel{{:}{=}}\mp Q,
\end{equation}
where the
sign corresponds to the parity preserving (`$-$') or reversing (`$+$') time~$t$
and~$\bar{t}$, respectively. Each nonlocality thus makes the conserved
current trivial; the new variables can be bosonic or fermionic. Hence,
starting with an equation~$\cE$, one calculates several conserved
currents for it and \emph{trivializes} them by introducing a
\emph{layer} of nonlocalities whose derivatives are still local
differential functions. This way the number of fields is increased and
the system is extended by new substitution rules. Moreover, it may
acquire new conserved currents that depend on the nonlocalities and
thus specify the second layer of nonlocal variables with nonlocal
derivatives. Clearly, the procedure is self\/-\/reproducing.
An example of fixing the derivatives of a nonlocal variable
is given in the text output. 
So, one keeps computing conserved currents and adding the layers of nonlinearities until an extended system~$\smash{\tilde{\cE}}$ is achieved such that its linearization~$\Lin\smash{\tilde{\cE}}$ has a (shadow of a, \cite{JKKersten}) symmetry~$\cR$; this
 symmetry of $\Lin\smash{\tilde{\cE}}$ is precisely the resursion for the extended system~$\smash{\tilde{\cE}}$.

The superscript `layers' (if non\/-\/empty) in the symbol
${}_{\text{ord}}^{\text{layers}}\cR_{\text{weight}}^{\sharp}$
indicates the required number of layers of the nonlocal variables
assigned to conserved currents. The symbol `$\sharp$' denotes the number of recursions found for a given differential order,
weight, and given nonlocalities.
In the sequel, we denote local recursion operators by $L$ and
nonlocal 
ones by $N$. The symbol $Z$ is used to denote
a nilpotent recursion operator whose powers are
equal zero except for a finite set of them, and $\varSigma$ is a
super\/-\/recursion that swaps the parities of the flows.

\begin{table}[tbph]
\caption{The classification of coupled super\/-systems
and their recursions with respect to
the primary weights $|f|=|b|=|D_t|=|D_{\bar{t}}|=\oh$.}
\label{Tab1}
\centerline{
\begin{tabular}{|
  l | 
  c |}
\hline
\eqref{DoubleLayer}\quad
$\left\{\begin{aligned}f_t&=\sd{b}+fb,\\ b_t&=\sd{f}\end{aligned}
\right.$ &
${}_0^2 N_{-1\frac{1}{2}}^1$,\ ${}_{\frac{1}{2}}^2 N_{-2}^1$,\
${}_{\frac{1}{2}}^2 N_{-2\frac{1}{2}}^1$,\ ${}_{\frac{1}{2}}^2 N_{-3}^1$ \\
\hline
\eqref{BurgSystem}\quad
$\left\{ \begin{aligned}f_t&=\sd{b},\\ b_t&=\sd{f}+b^2 \end{aligned}
\right.$ &   ${}_1^1 N_{-1}^1$ \\
\hline
\eqref{Quad}\quad
$\left\{\begin{aligned}f_t&=-\alpha\,fb,\\ b_t&=\sd{f}+b^2\end{aligned}
\right.$ &
$\begin{aligned}
\alpha&=2:\quad
 {}_{\frac{1}{2}} L_{-2}^{1},\
{}_{\frac{1}{2}} L_{-2\frac{1}{2}}^1,\ {}_{0}^0 Z_{-2}^1,\ 
{}_{0}^0 \varSigma_{-2}^1,\ {}_{0}^0 \varSigma_{-2\frac{1}{2}}^1,\ %
{}_{0}^0 \varSigma_{-2\frac{1}{2}}^1;\phantom{\Bigl(} \\
\alpha&=1:\quad{}
{{}_{0}^0 Z_{-2\frac{1}{2}}^1,\ {}_{\frac{1}{2}} Z_{-3}^1};
\qquad{\alpha=4:\quad
{}_{1} L_{-3\frac{1}{2}}^1}\phantom{\Bigl(}\\
\beta&\mathrel{{:}{=}}-1/\alpha:\quad
\beta\in\left\{-\tfrac{3}{4},-\tfrac{1}{3},
-\tfrac{1}{6},-\tfrac{1}{8},1,\pm\tfrac{3}{2},\pm2,\tfrac{5}{2},3\right\}
\end{aligned}$
\\
\hline
\eqref{stPar}\quad
$\left\{\begin{aligned}
  f_{\bar{t}}&=\sd{f}+b^2,\\
  b_{\bar{t}}&=\sd{b}+fb\end{aligned}
\right.$ &
${}_0 \varSigma_{-\frac{1}{2}}^1,\ {}_{\frac{1}{2}}L_{-1}^1,\ %
 {}_{\frac{1}{2}}\varSigma_{-\frac{3}{2}}^1$
\\
\hline
\end{tabular}%
}
\end{table}

It turns out that the equations presented in Table~\ref{Tab1} exhibit
practically the whole variety of properties that super\/-\/PDE of
mathematical physics possess. Let us discuss them in more detail.


\section{A fermionic extension of the Burgers equation}\label{SecExt}
The system 
\begin{equation}\label{DoubleLayer}
f_t=\sd{b}+fb,\qquad    b_t=\sd{f}
\end{equation}
is homogeneous w.r.t.\ a unique set of weights
$|f|=|b|=\oh$, $|D_t|=\oh$, $|D_x|=1$.
System~\eqref{DoubleLayer} admits
symmetries $(f_s$, $b_s)$ for all weights~$|D_s|\geq \oh$. 

We recognize that system~\eqref{DoubleLayer} is related to the
fermionic extension of the Burgers equation~\cite{Kiev2005}. Namely,
consider the fermionic super\/-\/field~$\phi=\eta(t,x)+\theta\xi(t,x)$
of weight~$0$ whose derivatives are $\phi_t=f$, $\sd{\phi}=b$
and which thus potentiates the second equation in~\eqref{DoubleLayer}.
Then from the first equation in~\eqref{DoubleLayer} we get the scalar equation $\phi_{tt}=\phi_x+\phi_t\cdot\sd{\phi}$. Its component form is the evolutionary system
\begin{equation}\label{WithAlpha}
\eta_x=\eta_{tt}-\xi\eta_t,\quad
\xi_x=\xi_{tt}-\xi\xi_t+\alpha\cdot\eta_t\eta_{tt},\qquad\alpha=1,
\end{equation}
where the variable~$x$ is the evolution parameter (the time) and the coordinate~$t$ is the new spatial variable.
System~\eqref{WithAlpha} is precisely the $\alpha=1$
fermionic {(super)}\/extension of the Burgers equation upon the
functions $b(t,x)$ and $w(t,x)$, see~\cite[Eq.~(16)]{Kiev2005},
rewritten by setting $\xi(t,x)=-b(x,t)$ and $\eta(t,x)=w(x,t)$.
Hence we conclude that the scalar super\/-\/field equation
$\phi_{tt}=\phi_x+\phi_t\cdot\sd{\phi}$ provides a one\/-\/component representation for that system;
we recall that a different one\/-\/component representation of the
fermionic extension for the Burgers equation derived in~\cite{Kiev2005}
is an evolution equation on an associative algebra.
The geometry of the fermionic (super-)\/extension~\eqref{WithAlpha}
was extensively studied
in~\cite{Kiev2005}; in particular, recursion operators were constructed
for its symmetries. We emphasize that the correlation between
system~\eqref{DoubleLayer} and the extended Burgers equation was not
noticed in~\cite{Dubna05}, where \eqref{DoubleLayer} was first
described.

The geometry of system~\eqref{DoubleLayer} itself is essentially nonlocal, which is illustrated by calculating the conserved currents using \SsTools, trivializing them, and constructing the recursion operators that involve the new nonlocalities.
It is very likely that system~\eqref{DoubleLayer} has only one conserved current which is already in use for specifying~$\phi$.
In the second layer, many nonlocal conservation laws and hence many new variables appear. This is discovered by \SsTools\ as follows.

The calculation of conservation laws for
evolutionary super\/-\/systems with homogeneous polynomial
right\/-\/hand sides is performed by using the procedure
\verb"ssconl":
\begin{verbatim}
 ssconl(N,tw,mincw,maxcw,afwlist,abwlist,pdes);
\end{verbatim}
where\\
\verb" N       "$\ldots$ the number of superfields $\theta^i$;\\
\verb" tw      "$\ldots$ $2$ times the weight $|D_t|$;\\
\verb" mincw   "$\ldots$ minimal weight of the conservation law;\\
\verb" maxcw   "$\ldots$ maximal weight of the conservation law;\\
\verb" afwlist "$\ldots$ list of weights of the fermionic fields \texttt{f(1)},$\ldots$,\texttt{f(nf)};\\
\verb" abwlist "$\ldots$ list of weights of the bosonic fields \texttt{b(1)},$\ldots$,\texttt{b(nb)};\\
\verb" pdes    "$\ldots$ \parbox[t]{12cm}{list of the equations for which a conservation law must be found.}

\smallskip\noindent%
The ansatz for the differential polynomial components of a conserved
current is composed in full generality (certainly, it has nothing to do
with the axioms for the symmetry flows).
Again, the global positive integer variable \verb"max_deg" determines the
highest power of a bosonic variable of nonpositive weight or a
fermionic variable of negative weight and all their
(super-)\/derivatives in any ansatz.
The procedure \verb"ssconl" (and \verb"wgts" and \verb"linearize" as well) is indifferent w.r.t.\ the presence of assignments \verb"=" and \verb"=>" in \verb"pdes".

The fact that the current is conserved on a given system
\verb"pdes" leads to the algebraic system for the undetermined
coefficients, which is further solved automatically by
\Crack~\cite{WolfCrack}. Having obtained a conserved current, we define
the new bosonic or fermionic dependent variable (the nonlocality) using
the standard rules~\eqref{EqNewNonlocal}.
An example illustrating the run of \verb"ssconl" is given in
the test output. 

So, in addition to the potential $\phi$, let us construct
the fermionic variable~$v$ whose
weight $|v|=\tfrac{3}{2}$ is minimal (other admissible nonlocal variables have greater weights):
we calculate conservation laws involving $\phi$ and then we set
$v_t=\sd{b}\cdot \phi fb+f_x \phi f$ and
$\sd{v}=-\sd{b}\cdot fb+\sd{f}\cdot\sd{b}\cdot\phi+b_x\phi f$.
Now there appear nontrivial solutions to the determining equations
for recursion operators.
By convention, the capital $V$ denotes the `linearized' counterpart of
$v$ and likewise $\Phi$ for~$\phi$. Recall that the correlation between
the `linearized' variables in the solutions~$\cR$ obtained by \SsTools\
and the recursion operators~$R$ that act on the symmetries is as
follows, see~\eqref{RecBurgers}: the local variables $F$, $B$, and
their derivatives denote the corresponding components of the flows, and
the linearized nonlocalities, e.g.\ $\Phi$, act by the rules that
follow from their derivation formulas; we thus have the interpretation
$\Phi={\mathcal{D}}^{-1}(B)$.

We obtain the recursion of
zero differential order with nonlocal coefficients: 
\[
\cR_{[-1\frac{1}{2}]} =
\binom{-\sd{b}\cdot \phi f B + \phi vF  + v\cdot B}%
      {\sd{b}\,\phi f F  - vF + v\phi \cdot B}.
\]
Also, we get another operator, which is nonlocal and has nonlocal coefficients, 
\[
\cR_{[-2]} =
\binom{\sd{b}\,V\phi -\sd{f}\,\sd{B}\,\phi f-\sd{f}\,\sd{b}\,\Phi f+\sd{f}\,\sd{b}\,F\phi
    +\sd{f}\,V+V\phi fb}%
 {\sd{B}\,\sd{b}\,\phi f+\sd{b}\,V-\sd{b}\,F\phi fb+\sd{f}\,\sd{b}\,V\phi f+\sd{f}\,V\phi -Vfb};
\]
it contains $\phi$ as well as $\Phi$ and~$V$.
The coefficients of the recursions $\cR_{[-5/2]}$ and $\cR_{[-3]}$
found for $|s_R|=-2\oh$ and $|s_R|=-3$ are also nonlocal.


\section{A super\/-\/field representation for the Burgers equation}
Consider the system 
\begin{equation}\label{BurgSystem}
f_t=\sd{b},\qquad b_t=\sd{f}+b^2,
\end{equation}
which has the unique set of weights
$|f|=|b|=\oh$, $|D_t|=\oh$, $|D_x|=1$.
Equation~\eqref{BurgSystem} admits the infinite
sequence~\eqref{SymBurg} of higher
symmetries $f_s=\phi^f$, $b_s=\phi^b$ at all
(half-)\/integer weights $|D_s|\geq \oh$.    
Also, there is another infinite sequence~\eqref{BurgSSym}
of symmetries for~\eqref{BurgSystem} at all (half-)integer weights
$|D_{\bar{s}}|\geq \oh$ of the odd `time'~$\bar{s}$.

System~\eqref{BurgSystem} is obviously related to the bosonic
super\/-\/field Burgers equation
\begin{equation}\label{BurgersInverse}
b_x=b_{tt}-2bb_t,\qquad b=b(x,t,\theta).
\end{equation}
We emphasize that the role of the independent coordinates $x$ and $t$
is reversed w.r.t.\ the standard interpretation of $t$ as
the time and $x$ as the spatial variable.
The Cole\/--\/Hopf substitution $b=-u^{-1}u_t$ from the heat equation
$u_x=u_{tt}$
provides the solution for the bosonic component of~\eqref{BurgSystem}.

Let us introduce the bosonic nonlocality $w(x,t,\theta)$
of weight $|w|=0$ by trivializing the conserved form of the first equation in~\eqref{BurgSystem}. Namely, we specify the derivatives
of~$w$ by setting
\[
\sd{w}=-f,\quad  w_t=-b.
\]
Note that the variable $w$ is a potential for both fields $f$ and $b$.
The nonlocality $w$ satisfies the potential Burgers equation
$w_x=w_{tt}+w_t^2$ such that the formula $w=\ln u$ gives the solution;
the relation $f=-\sd{w}$ determines the fermionic component in
system~\eqref{BurgSystem}.

Now we extend the set of dependent variables $f$, $b$, and $w$ by the
symmetry generators
$F$, $B$, and $W$ that satisfy the respective linearized relations,
\[
F_t=\sd{B},\quad B_t=\sd{F}+2bB,\quad \sd{W}=-F,\quad W_t=-B.
\]
The linearization correspondence between the fields is $f\mapsto F$, $b\mapsto B$, and $w\mapsto W$.
In this setting, we obtain the recursion of weight $|s_R|=-1$:
\begin{equation}\label{RecBurgers}
\cR_{[-1]}=\binom{F_x-\sd{f}\,F+f_x\,W}%
                        {B_x-\sd{f}\,B+b_x\,W}\ \Longleftrightarrow\ %
R=\begin{pmatrix}
  D_x-\sd{f}+f_x\,\sd^{-1} & 0 \\
  b_x\,\sd^{-1} & D_x-\sd{f}
\end{pmatrix}.
\end{equation}
The method for constructing $\cR_{[-1]}$ is illustrated in the
test output. 


The recursion $\cR_{[-1]}$ 
provides two sequences of higher
symmetries for system~\eqref{BurgSystem}:
\begin{equation}\label{SymBurg}
\binom{f_t}{b_t}\mapsto
\binom{\sd{b_x}-\sd{f}\sd{b}-f_xb}{\sd{f_x}-(\sd{f})^2-b^2\sd{f}+bb_x}
   \mapsto\cdots,\ 
\binom{f_x}{b_x}\mapsto
\binom{f_{xx}-2\sd{f}f_x}{b_{xx}-2\sd{f}b_x}\mapsto\cdots.
\end{equation}
The same recursion 
proliferates two experimentally found first order flows to
two infinite sequences
of symmetries with the odd parameters~$\bar{s}$:
\begin{equation}\label{BurgSSym}
\binom{\sd{f}}{\sd{b}} \mapsto
\binom{\sd{f_x}-(\sd{f})^2-f_xf}{\sd{b_x}-\sd{f}\,\sd{b}-b_xf}
\mapsto\cdots,\ 
\binom{f\sd{b}-b\,\sd{f}+b_x}{b\sd{b}-f\,\sd{f}+f_x-fb^2}
\mapsto\cdots.
%
\end{equation}

\begin{rem}\label{ProfitInverse}\upshape
System~\eqref{BurgSystem} is not a supersymmetric extension
of~\eqref{BurgersInverse}; it is a representation of
the bosonic super\/-\/field Burgers equation.
The flows in~\eqref{SymBurg} become purely bosonic
in the coordinates $\sd{f}$, $b$.
The standard recursion $R=D_t+\tfrac{1}{2}b+\tfrac{1}{2}b_tD_t^{-1}$
for the Burgers equation acts `across' the two sequences
in~\eqref{SymBurg} and maps $(f_t, b_t)\mapsto(f_x, b_x)$;
again, we note that the independent coordinates
are reversed in~\eqref{BurgersInverse}.

However, from the above reasonings we gain
two sequences of symmetries~\eqref{BurgSSym},
which are \emph{not} reduced to the
bosonic $(x,t)$-independent symmetries of the Burgers equation.
We finally recall that the Burgers equation~\eqref{BurgersInverse} has
infinitely many higher symmetries \cite{JKVin} that depend explicitly on the base coordinates $x$, $t$ and thus exceed the set of axioms on
p.~\pageref{pAxioms}.
\end{rem}

\section{A multiplet of super\/-\/systems}
In this section we consider the systems
\begin{equation}\label{Quad}
f_t=-\alpha fb,\qquad b_t=\sd{f}+b^2.
\end{equation}
Here the normal growth equation upon
\[
f(t,x;\theta)=f(0,x;\theta)\cdot\exp\left(
   -\alpha\int_0^t b(\tau,x;\theta)\,\Id\tau\right)
\]
is coupled with the perturbed blow-up equation upon $b(t,x;\theta)$.
System~\eqref{Quad} has multiple homogeneities, and we
let the tuple $|f|=|b|=\oh$, $|D_t|=\oh$, $|D_x|=1$ be the primary
`reference system.' The symmetry integrable cases differ by the values of the parameter~$\alpha$. We discover that for $\alpha=1$, $2$, and $4$ systems~\eqref{Quad} admit the symmetries with both even and odd times $s$, $\bar{s}$. The equations in this
triplet demonstrate different 
properties. The geometry of the $\alpha=2$ system
is quite extensive: this system admits a continuous sequence of
symmetries for all (half\/-)\/integer weights $|D_s|\geq \oh$,
four local recursions (one is nilpotent), and three local
super\/-\/recursions. The equation for $\alpha=1$ admits fewer
structures, and the case $\alpha=4$ for~\eqref{Quad} is rather
poor. All the three equations possess higher symmetries with
parity reversing
times~$\bar{s}$. In what follows, we indicate the presence of
recursion operators for these systems in certain weights $|s_R|$,
$|\bar{s}_R|$.

Next, we find values of the parameter $\beta=-1/\alpha$ such
that the respective systems~\eqref{Quad} possess symmetries with
parity reversing
times~$\bar{s}$ but have no higher commuting flows with the 
times~$s$. Eleven cases are then realized and we specify the
admissible weights~$|\bar{s}|$ for each~$\beta$.

\subsection{Case $\alpha=2$.}
First we fix $\alpha=2$ 
and consider~\eqref{Quad}: we get
$ 
f_t=-2fb$, $b_t=\sd{f}+b^2$.
The weights for symmetries are $|D_s|=\oh$, $|D_s|=1$, and then
equation~\eqref{Quad} admits a continuous chain of symmetry flows
for all (half-)integer weights $|D_s|\geq 2\oh$.  
Surprisingly, no nonlocalities are needed to construct the recursion
operators, although there are many conservation laws for this system.
Hence we obtain purely local recursion operators~$\cR$ that
propagate the symmetries:
$\vph=(F,B)\mapsto\vph'=\cR$ for any $\vph$. We get
an operator of weight $-3\oh$,
\begin{equation}\label{BreadthRec}    
\cR_{[-7/2]}=\binom{0}%
  {(\sd{f})^3fF+6(\sd{f})^2fb^2F+12\sd{f}\cdot fb^4F+8fb^6F},
\end{equation}
which is triangular since $\cR^b$ does not contain~$B$ and, moreover, which is nilpotent.
The above recursion is a recurrence relation~\cite{Kiev2005}
that is well\/-\/defined for all symmetries of~\eqref{Quad}.
The recursion~$\cR_{[-2]}$
of weight $|s_R|=-2$ is also triangular; we have
\[          
\cR_{[-2]} =
\begin{pmatrix}
\tfrac{11}{2}\sd{F}\,\sd{f}\,f + 11\sd{F}\,fb^2 + \tfrac{3}{2}(\sd{f})^2 F +
3\sd{f}\,Fb^2 + \tfrac{1}{2}f_xFf \\
  \begin{gathered}[t]  {} \\
   11\sd{B}\,fb^2 + 8\sd{b}\,Fb^2 + 22\sd{b}\,fBb + 7(\sd{f})^2 B +{}\\
   14\sd{f}\,Bb^2 + \tfrac{11}{2}\sd{f}\,\sd{B}\,f +
   \tfrac{5}{2}\sd{f}\,\sd{b}\,F +
   \tfrac{1}{2}b_xFf + f_x F b + 5 f_xfB
  \end{gathered}
\end{pmatrix},
\]
Further, we obtain the recursion~$\cR_{[-5/2]}$
of weight~$2\oh$; its components are
\begin{align*}  
\cR_{[-2\oh]}^f &=
  - 2\sd{b}\,Ffb^2 - \sd{F}\,\sd{f}\,fb - \sd{F}\,fb^3 - \tfrac{1}{2}f_xFfb
  - 2\sd{f}\,fBb^2, \\
\smash{\cR_{[-2\oh]}^b} &=
  \sd{B}\,fb^3 + \sd{b}\,Fb^3 + \sd{b}\,fBb^2 + \tfrac{1}{8}\sd{f_x}Ff +{}\\
 &\quad+ \tfrac{1}{2}\sd{F}b^4 + \tfrac{1}{2}\sd{F}(\sd{f})^2
  + \sd{F}\,\sd{f}\,b^2 + \tfrac{1}{8}\sd{F}\,f_xf + (\sd{f})^2Bb +{}\\
 &\quad+ \sd{f}\,Bb^3 + \sd{f}\,\sd{B}\,fb + \sd{f}\,\sd{b}\,Fb
  + \sd{f}\,\sd{b}\,fB + \tfrac{3}{8}\sd{f}\,F_xf + {}\\
 &\quad+\tfrac{1}{4}\sd{f}\,f_xF
  + \tfrac{1}{2}b_xFfb + \tfrac{1}{2}F_xfb^2 + \tfrac{1}{4}f_xFb^2
  + \tfrac{1}{2}f_xfBb.
\end{align*}
The local recursion with $|s_R|=-3$ is huge. 


For $\alpha=2$, system~(\ref{Quad})
admits at least three super\/-\/recursions
${}^t(R^f$, $R^b)$ such that the parities of $R^f$ and
$R^b$ are opposite to the odd parity for $f$ (and hence for $F$) and to
the even parity of $b$ and~$B$. This
is possible owing to the presence of the 
variable~$\bar{s}_R$.
The triangular zero\/-\/order super\/-\/recursions are 
${\bar{\cR}}_{[-2]}^f =   
{4\sd{f}\,Ffb+8Ffb^3}$, ${\bar{\cR}}_{[-2]}^b=
-4\sd{b}\,Ffb+2(\sd{f})^2F+6\sd{f}\,Fb^2+4\sd{f}\,fBb$
${}-f_xFf+4Fb^4+8fBb^3$
and   
\[
{\bar{\cR}}_{[-2\frac{1}{2}]} =
\binom{-\sd{f}\,f_xF-2f_xFb^2}%
{\sd{b}\,f_xF-\sd{f}\,b_xF+\sd{f}\,f_xB-2b_xFb^2+2f_xBb^2}
\]
for the weights $|\bar{s}_R|=-2$ and $|\bar{s}_R|=-2\oh$, respectively;
the third super\/-\/recursion found
for $|\bar{s}_R|=-2\oh$ is very large.       
Quite naturally, system~\eqref{Quad} has infinitely many
$\bar{s}$-symmetries if~$\alpha=2$.

\subsection{Case $\alpha=1$.}
Setting $\alpha=1$ in~\eqref{Quad}, we obtain the system 
$ 
f_t=-fb$, $b_t=\sd{f}+b^2$.
The default set of weights is again
$|f|=|b|=\oh$, $|D_t|=\oh$, and $|D_x|=1$.
The sequence of symmetries is not continuous
and starts later than for the chain in the case $\alpha=2$.
We find out that there are symmetry flows
if either $|D_s|=|D_t|=\oh$ (the equation itself),
$|D_s|=|D_x|=1$ (the translation along~$x$),
or $|D_s|\geq 3\oh$ such that a
continuous chain starts for all (half-)\/in\-te\-ger
weights~$|D_s|$. 

Similarly to the previous case, no nonlocalities are needed to
construct the recursions, which therefore are purely local.
The recursion operator   
$\cR_{[-5/2]}^f=0$, $\cR_{[-5/2]}^b=
{(\sd{f})^2\,Ff+3\sd{f}\,Ffb^2+\tfrac{9}{4}Ffb^4}$
of maximal weight $|s_R|=-2\oh$ is nilpotent: $\cR^2=0$.
For the succeeding weight $|s_R|=-3$,
we obtain a nilpotent local recursion~$\cR_{[-3]}$
whose components are given through
\begin{align*} 
\cR_{[-3]}^f&=
  \tfrac{5}{3}\sd{F}\,(\sd{f})^2f+\tfrac{5}{2}\sd{F}\,\sd{f}\,fb^2
  -\tfrac{5}{3}(\sd{f})^3F-\tfrac{5}{2}(\sd{f})^2Fb^2 +{}\\
 &+{5}\sd{f}\,\sd{b}\,Ffb+\tfrac{20}{3}\sd{f}\,f_xFf
  +\tfrac{15}{2}f_xFfb^2,\\
\cR_{[-3]}^b&=
  \sd{f_x}\,Ffb-\tfrac{105}{2}\sd{F}\,\sd{b}\,fb^2
  -\tfrac{160}{3}\sd{F}\,\sd{f}\,\sd{b}\,f+{11}\sd{F}\,f_xfb+{}\\
 &+\tfrac{5}{3}(\sd{f})^2\sd{B}f+\tfrac{5}{3}(\sd{f})^2\sd{b}\,F
  +\tfrac{5}{2}\sd{f}\,\sd{B}\,fb^2+\tfrac{5}{2}\sd{f}\,\sd{b}\,Fb^2-{}\\
 &-{55}\sd{f}\,\sd{b}\,fBb+\tfrac{17}{3}\sd{f}\,b_xFf
  +\sd{f}\,f_xfB+\tfrac{23}{2}b_xFfb^2+\tfrac{183}{2}f_xfBb^2.
\end{align*}
The differential order of $\cR_{[-3]}$ is positive.

\subsection{Case $\alpha=4$.}
Finally, let $\alpha=4$; then system~\eqref{Quad}
acquires the form   
$ 
f_t=-4fb$, $b_t=b^2+\sd{f}.$
Again, the primary set of weights is
$|f|=|b|=\oh$, $|D_t|=\oh$, $|D_x|=1$.
System~\eqref{Quad}
admits the symmetries $(f_s$, $b_s)$ of weights
$|D_s|=\oh$, $1$ or $|D_s|\geq 3\oh$ w.r.t.\ the primary set above.
This situation coincides with the case~$\alpha=1$.
Again, no nonlocalities are needed for constructing the recursion
$\cR_{[-7/2]}$ of weight $|s_R|=-3\oh$;
this operator is relatively big:
\begin{align*} 
\smash{\cR_{[-3\oh]}^f}&=
   -12\sd{b}\,Ffb^4-\sd{F}\,(\sd{f})^2fb
   -{4}\sd{F}\,\sd{f}\,fb^3 - 3\sd{F}\,fb^5-{}\\
  &-{4}(\sd{f})^2fBb^2-{4}\sd{f}\,\sd{b}\,Ffb^2
   -\tfrac{2}{3}\sd{f}\,f_xFfb-12\sd{f}\,fBb^4-{2}f_xFfb^3,\\
\smash{\cR_{[-3\oh]}^b}&=
   3\sd{B}fb^5+3\sd{b}\,Fb^5+9\sd{b}\,fBb^4
   +\tfrac{1}{9}\sd{f_x}\,\sd{f}\,Ff-\tfrac{1}{3}\sd{f_x}\,Ffb^2+{}\\
  &+\sd{F}\,\sd{b}\,fb^3
   +\tfrac{1}{4}\sd{F}\,(\sd{f})^3+\tfrac{5}{4}\sd{F}\,(\sd{f})^2b^2
   +\tfrac{7}{4}\sd{F}\,\sd{f}\,b^4+{}\\
  &+\tfrac{5}{18}\sd{F}\,\sd{f}\,f_xf
   +\tfrac{1}{2}\sd{F}\,f_xfb^2+(\sd{f})^3Bb
   +{4}(\sd{f})^2Bb^3+{}\\
  &+(\sd{f})^2\sd{b}\,Fb+(\sd{f})^2\sd{b}\,fB
   +\tfrac{2}{9}(\sd{f})^2F_xf+\tfrac{1}{6}(\sd{f})^2f_xF+{}\\
  &+{4}\sd{f}\,\sd{B}\,fb^3
   +{4}\sd{f}\,\sd{b}\,Fb^3 + {10}\sd{f}\,\sd{b}\,fBb^2
   +\sd{f}\,F_xfb^2+{}\\
  &+\tfrac{2}{3}\sd{f}\,f_xFb^2
   +\tfrac{5}{3}\sd{f}\,f_xfBb
   +{2}b_xFfb^3+F_xfb^4+\tfrac{1}{2}f_xFb^4+f_xfBb^3+{}\\
  &+\tfrac{3}{4}\sd{F}\,b^6+\sd{F}\,\sd{f}\,\sd{b}\,fb
   +(\sd{f})^2\sd{B}\,fb+3\sd{f}\,Bb^5+\tfrac{2}{3}\sd{f}\,b_xFfb.
\end{align*}
No nilpotent recursion operators were found for system~(\ref{Quad})
if~$\alpha=4$.

\begin{rem}\label{InfByNilpotent}\upshape
We observe that an essential part of recursion operators for
supersymmetric PDE are nilpotent.
At present, it is not clear how the nilpotent recursion operators
contribute to the integrability of supersymmetric systems and what
invariants they describe or symptomize.
We emphasize that this property does not always originate from
the rule `$f\cdot f=0$', but this is an immanent feature of the
symmetry algebras.
\end{rem}


\begin{prob}\upshape
Construct an equation~$\cE$ that admits
nilpotent differential recursion operators $\{R_1,\ldots\mid
R_i^{n_i}=0\}$ which generate an infinite sequence of symmetries
$\vph$, $R_{i_1}(\vph)$, $R_{i_2}\circ R_{i_1}(\vph)$, $\ldots$
for~$\cE$. Here we assume that
at least two operators (without loss of generality, $R_1$ and $R_2$)
do not commute and hence the flows never become zero.
\end{prob}

\subsection{The systems with parity reversing parameters~$\bar{s}$}
Now we consider eleven systems~\eqref{Quad} that admit symmetries with
parity reversing parameters~$\bar{s}$ but do not possess any commuting
$s$-\/flows except for the cases $s=t$, $|D_t|=\oh$, and $s=x$,
$|D_x|\equiv 1$. These systems are enumerated by the
parameter~$\beta=-1/\alpha$ whose usage makes the ordering simpler: we
conjecture that the systems possessing higher $\bar{s}$-\/symmetries
constitute an infinite family corresponding, in particular, to
(half-)\/integer positive~$\beta$'s.
In Table~\ref{Tab2} we show the numbers of symmetries with a certain
weight~$|\bar{s}|$ for each of the systems~\eqref{Quad} specified by~$\beta$. For convenience, we use the computer notation for the weights, that is, we multiply them by~$2$. Empty boxes correspond to no symmetries in that particular weight.

\begin{table}[tbph]
\caption{The symmetry structure for special systems~\eqref{Quad}.}
\label{Tab2}
\centerline{
\begin{tabular}{|
  c | 
  c | c | c | c | c | c | c | c | c |}
\hline
$\beta\downarrow{}\ \mid\ 2|D_{\bar{s}}|\to{\mathstrut}$
   &2&3&4&5&6&7&8&9&10\\
\hline
$1$ & 
   $\bullet$ & $\bullet$ & & & & & & & \\
\hline
$3/2$ & 
   & & $\bullet$ & & & & $\bullet$ & & \\
\hline
$2$ & 
   & & & & $\bullet$ & & & & $\bullet$ \\
\hline
$-3/2$ & 
   & & & & & & $\bullet$ & $\bullet\bullet\bullet$ &
                                    $\bullet\bullet\bullet$ \\
\hline
$-1/6$ & 
   & & & & & & $\bullet$ & $\bullet$\,$\bullet$ & $\bullet$ \\
\hline
$5/2$ & 
   & & & & & & $\bullet$ & & \\
\hline
$-2$ & 
   & & & & & & & & $\bullet$ \\
\hline
$-3/4$ & 
   & & & & & & & & $\bullet$ \\
\hline
$-1/3$ & 
   & & & & & & & & $\bullet$ \\
\hline
$-1/8$ & 
   & & & & & & & & $\bullet$ \\
\hline
$3$ & 
   & & & & & & & & $\bullet$ \\
\hline
\end{tabular}
}
\end{table}


\section{A system with parity reversing times}
Let the time~$\bar{t}$ and the parameters~$\bar{s}$ be parity
reversing. Then there is a unique system that satisfies the axioms and the weight assumptions $|f|=|b|=|D_{\bar{t}}|=\oh$,
\begin{equation}\label{stPar}
  f_{\bar{t}}=\sd{f}+b^2,\quad
  b_{\bar{t}}=\sd{b}+fb.
\end{equation}
Clearly, the weights are uniquely defined in~\eqref{stPar}.
The numbers of symmetries for system~\eqref{stPar}
are arranged in Table~\ref{TabSTPar};
note that the integer weights correspond to~$|s|$ and the
half\/-\/integer values stand for~$|\bar{s}|$.
\begin{table}[tbph]
\caption{The number of symmetries for system~\eqref{stPar}.}
\label{TabSTPar}
\centerline{
\begin{tabular}{|
  l | 
  c | c | c | c | c |}
\hline
$|D_s|\in\BBZ,\ |D_{\bar{s}}|\in\BBZ+\oh\phantom{\Bigr)}$ &
  $\oh$ & $1$ & $1\oh$ & $2$ & $2\oh$ \\
\hline
$\sharp\sym$      & $2$    & $4$  & $2$     & $1$  & $1\mathstrut$\\
\hline
\end{tabular}
}
\end{table}

The flows commuting with~\eqref{stPar} are proliferated by three local
recursion operators. The nilpotent super\/-\/recursion~$\cR_{[-1/2]}$
of weight~$|\bar{s}_R|=-\oh$ and
the local recursion with $|s_R|=-1$ are given through
\begin{equation}\label{RecTparSpar}
\cR_{[-\frac{1}{2}]}=\binom{fF}{\tfrac{1}{2}bF-fB},\qquad
\cR_{[-1]}=\binom{b\cdot\sd{B}-\sd{b}\cdot B-\sd{f}\cdot F}%
   {\tfrac{1}{2}b\cdot\sd{F}-\sd{f}\cdot B}.
\end{equation}
Another parity\/-\/reversing recursion for system~\eqref{stPar} is
\[
\cR_{[-1\frac{1}{2}]}=\binom{f\sd{b}\cdot B+f\sd{f}\cdot F-fb\cdot\sd{B}}%
{\tfrac{1}{2}b\sd{b}\cdot B+\tfrac{1}{2}bf\cdot\sd{F}-\tfrac{1}{2}b^2\cdot\sd{B}+
    \tfrac{1}{2}b\sd{f}\cdot F-f\sd{f}\cdot B},
\]
here we have~$|\bar{s}_R|=\smash{-1\tfrac{1}{2}}$.

In the test output, 
we present the \SsTools\ calls that demonstrate how the recursion
$\cR_{[-1/2]}$ is obtained and we explain how the output is translated
into the geometric language.

\section{Conclusion}
Using the presented computational techniques,
we discovered that four classes of nonlinear coupled boson\/-\/fermion
systems satisfy the set of axioms, which are believed natural in view
of the nontriviality and nondegeneracy they ensure.
Two of the classes are related to the Burgers equation,
which is one of the most relevant and model\/-\/like equations
of mathematical physics. Namely, they are a fermionic
(super-)\/extension and a supersymmetric representation of the Burgers
equation, respectively. The latter admits symmetries such that both the
bosonic and fermionic fields are involved and such that the flows do
not retract to the standard, purely commutative setting. Also, we found
that a unique evolutionary system in this set of homogeneity weights
has the parity reversing time~$\bar{t}$.
Finally, we obtained a multiplet of coupled normal evolution and
explosion systems; the multiplet seems to be essentially infinite
containing all (half)\/integer values $\beta\in\oh\BBN$ of the
parameter at the least. Only three of the systems from this
multiplet corresponding to $\alpha\in\{1,2,4\}$ admit higher symmetry
flows with the parity preserving parameters~$s$; the parameters
$\bar{s}$ for higher symmetries of other systems are parity reversing.
We do not know whether these three systems are distinguished by any
relation to the three classical sequences of the complex
semi\/-\/simple Lie algebras.
Hence we see that the suggested classification scheme yields both
generalizations of known integrable systems and also new models, which
are expected to be relevant in nature in view of their vast symmetry
properties: recursion operators were constructed for the above systems.

The search for super\/-\/systems together with their symmetries and
further revealing the recursion operators were performed by the
\SsTools\ package, which was used in two modes, respectively. The
breadth first search method for the recursions, see
Remark~\ref{Breadth}, which is valid for the systems that admit
multiple homogeneities, proved to be a serious help in debugging
\SsTools. Thus we conclude that the general classification problem (see
section~\ref{SecCommFlows}), which is an immense task of modern computer
physics, is becoming tractable with this computer algebra package.

\subsection*{Acknowledgements}
The authors thank V.\,V.\,Sokolov and A.\,S.\,Sorin for helpful
discussions and are grateful to the referee for remarks and
suggestions.
In addition, T.\,W.\ thanks W.\,Neun for discussions and
the SHARCNET consortium \texttt{www.sharcnet.ca} for computer access. The research of A.\,K.\ was
partially supported by Brock University under contract 609-607-831.
A part of this research was done while A.\,K.\ was visiting at
Max Planck Institute for Mathematics (Bonn) and at Brock
University. Support from the IMA Workshop
`Symmetries and overdetermined systems of PDE' at University of
Minnesota is gratefully acknowledged.

\noindent\textbf{TEST RUN OUTPUT}   

\noindent%
Here we illustrate the process of
computing symmetries, recursions, and conserved currents by \SsTools.
First we consider the weights $|f|=|b|=|D_t|=\tfrac{1}{2}$ and find all
systems that satisfy these homogeneity weights and admit higher
symmetries. We discover that
system~\eqref{stPar} is a unique solution obtained by the call
\begin{verbatim}
  ssym(1,1,sw,{1},{1},{},{},{},{tpar,spar});
\end{verbatim}
where \verb"sw"${}\in\{1,2,3,4,5\}$ and the flag \verb"spar" is omitted whenever \verb"sw" is even. The weights for this system are uniquely defined, which is confirmed by the call
\begin{verbatim}
  wgts({df(f(1),t)=b(1)**2+d(1,f(1)),
        df(b(1),t)=f(1)*b(1)+d(1,b(1))},{1},{1},maxwt);
\end{verbatim}
here \verb"maxwt" can be any positive integer.

Now we present the intermediate computations that result in the recursion $\cR_{[-1/2]}$ for~\eqref{stPar}, see~\eqref{RecTparSpar}.
We compute the linearization of system~\eqref{stPar},
see section~\ref{pCalcRec} for its definition:
\begin{verbatim}
  linearize({df(f(1),t)=b(1)**2+d(1,f(1)),
             df(b(1),t)=f(1)*b(1)+d(1,b(1))},1,1);
\end{verbatim}
The output contains system~\eqref{stPar} and its linearization:
\begin{verbatim}
  df(f(2),t)=2*b(2)*b(1) + d(1,f(2));
  df(b(2),t)=d(1,b(2)) + f(2)*b(1) + f(1)*b(2);
\end{verbatim}
The linearization correspondence between the fields is $f=\mathtt{f(1)}\mapsto F=\mathtt{f(2)}$ and $b=\mathtt{b(1)}\mapsto B=\mathtt{b(2)}$; the scaling weights of $F={}$\verb"f(2)" and $B={}$\verb"b(2)" are equal to the weights of $f={}$\verb"f(1)" and
 $b={}$\verb"b(1)". Note that the numbers \verb"nf" and \verb"nb" of fields are automatically doubled. The linearized system is calculated at most once in any run of \SsTools; the output is pasted into the next call as is.

Now is the time to find the recursion such that $|\bar{s}_R|=-\oh$ and hence \verb"sw"$(=-2|\bar{s}_R|)=1$.
The reason why the substitutions \verb"=>" are used in the original equations is explained in Remark~\ref{RemSubst} on p.~\pageref{RemSubst}: by construction, the recursions are symmetries of the linearized system, and the original equations are used for
 intermediate substitutions only. We thus have
\begin{verbatim}
  ssym(1,1,1,{1,1},{1,1},
    {df(f(1),t)=>b(1)**2+d(1,f(1)),
     df(b(1),t)=>f(1)*b(1)+d(1,b(1)),
     df(f(2),t)=2*b(2)*b(1) + d(1,f(2)),
     df(b(2),t)=d(1,b(2)) + f(2)*b(1) + f(1)*b(2)},
    {},{},{tpar,spar,lin});
\end{verbatim}
The output contains a unique solution:
\begin{verbatim}
  1 solution was found.
    df(f(2),s)=f(2)*f(1);
    df(b(2),s)= - 1/2*f(2)*b(1) + f(1)*b(2);
\end{verbatim}
This is precisely the first recursion in~\eqref{RecTparSpar} for
system~\eqref{stPar}. The result is achieved by the intermediate call
of \Crack~\cite{WolfCrack} from \verb"ssym"; the solver can be handled
in full automatic mode by typing \verb"g 1000".

Let us present a similar calculation that yields the local recursion~\eqref{BreadthRec} for system~\eqref{Quad} with $\alpha=2$; note that this time the weights are not uniquely defined and we thus have the multiple homogeneity case,
see Remark~\ref{Breadth}.
The input is
\begin{verbatim}
  hom_wei:={{16,{3,3},{2,2}}};
  ssym(1,1,7,{1,1},{1,1},
    {df(f(1),t)=>-2*f(1)*b(1),
     df(b(1),t)=>d(1,f(1))+b(1)**2,
     df(f(2),t)=-2*f(2)*b(1)-2*f(1)*b(2),
     df(b(2),t)=d(1,f(2))+2*b(1)*b(2)},
    {},{},{lin,filter});
\end{verbatim}
the output is the recursion~\eqref{BreadthRec}:
\begin{verbatim}
  1 solution was found.
    df(f(2),s)=0;
    df(b(2),s)=d(1,f(1))**3*f(2)*f(1)
      + 6*d(1,f(1))**2*f(2)*f(1)*b(1)**2
      + 12*d(1,f(1))*f(2)*f(1)*b(1)**4 + 8*f(2)*f(1)*b(1)**6;
\end{verbatim}
Note that the admissible set of weights $|b|=|B|=0$, $|f|=|F|=-\oh$ for system~\eqref{Quad} and the value $|s_R''|=1$ would require the upper bound \verb"max_deg:=6;" for this solution.

By shifting the primary weight \verb"sw" and/or the additional
values \verb"sw_i" in \verb"hom_wei", one can inspect how new recursion operators appear and vanish.

\medskip
A computation of a conserved current using \verb"ssconl" is done as follows; here we get the conserved current for system~\eqref{DoubleLayer} having already extended it by
the variable~$\phi={}$\verb+f(2)+, see p.~\pageref{WithAlpha}.
The input is
\begin{verbatim}
  ssconl(1,1,5,5,{1,0},{1},
    {df(f(1),t)=d(1,b(1))+f(1)*b(1),
     df(b(1),t)=d(1,f(1)),
     df(f(2),t)=>f(1),
      d(1,f(2))=>b(1)});
\end{verbatim}
Then the output produced by \SsTools\ is        
\begin{verbatim}
  NEXT: BOSONIC CONSERVATION LAWS OF WEIGHT 5
  >>>>> Non-trivial conservation law:
  Pt = Db*f(1)*b - Df(1)*Db*f(2) - b *f(2)*f(1)
                                    x

  Pd(1) =  Db*f(2)*f(1)*b + f(1) *f(2)*f(1)
                                x
\end{verbatim}
By trivializing the conservation relation $D_t(\mathtt{Pt})+\sd{(\mathtt{Pd(1)})}\doteq0$, we construct the fermionic variable~$v$ of weight~$|v|=\frac{3}{2}$, see section~\ref{SecExt}.


\begin{thebibliography}{99}

\bibitem{MiShSok}
\by{A. V. Mikhailov, A. B. Shabat, V. V. Sokolov},
The symmetry approach to classification of integrable equations.
\book{What is integrability?} 
Series in Nonlinear Dynamics, Springer, Berlin, 1991, 115-184, V.~E.~Zakharov (ed).

\bibitem{Carstea}
\by{A. S. Carstea}, Extension of the bilinear formalism to
supersymmetric KdV\/-\/type equations, \jour{Nonlinearity} \vol{13}
(2000), 1645--1656;
\by{A. S. Carstea, A. Ramani, B. Grammaticos}, Constructing the soliton
solutions for $N=1$ supersymmetric KdV hierarchy, \jour{Nonlinearity}
\vol{14} (2001), 1419--1423.

\bibitem{KuperSS}
\by{B. A. Kupershmidt}, \book{Elements of superintegrable systems.
Basic technique and results}, Reidel, Dordrecht, 1987.

\bibitem{Tsuchida}
\by{T. Tsuchida, T. Wolf}, Classification of polynomial integrable
systems of mixed scalar and vector evolution equations,
\jour{J.~Phys.\ A\textup{:} Math.\ Gen.} \vol{38} (2005), 7691--7733.

\bibitem{JKKersten}
\by{I. S. Krasil'shchik, P. H. M. Kersten},
\book{Symmetries and recursion operators for classical and
supersymmetric differential equations}, Kluwer Acad.\ Publ.,
Dordrecht etc., 2000.

\bibitem{Kiev2005}
\by{A. V. Kiselev, T. Wolf},
Supersymmetric representations and integrable fermionic extensions
of the Burgers and Boussinesq equations,
\jour{SIGMA~-- Symmetry, Integrability
and Geometry\textup{:} Methods and Applications} \vol{2} (2006),
no.30, 19~p.\ \texttt{arXiv:math-ph/0511071}

\bibitem{Olver}
\by{P.J.~Olver}, \book{Applications of Lie groups to differential
equations}, $2$nd ed.,  
Springer, Berlin, 1993.

\bibitem{Wintern}
\by{B. Champagne, W. Hereman, and P. Winternitz}, The computer calculation of Lie point symmetries of large systems of differential equations, \jour{Comput.\ Phys.\ Commun.} \vol{66} (1991), 319-340.

\bibitem{FPM}
\by{A. V. Kiselev}, Methods of geometry of differential equations
in analysis of integrable models of field theory,
\jour{J.~Math.\ Sci.} \vol{136} (2006) no.6 `Geometry of Integrable
Models', 4295--4377.\ \texttt{arXiv:nlin.SI/0406036}

\bibitem{Lstar}
\by{P. Kersten, I. Krasil'shchik, and A. Verbovetsky},
  Hamiltonian operators and $\ell^*$-coverings,
\jour{J.~Geom.\ Phys.} \vol{50} (2004) no.1-4, 273--302.

\bibitem{WHER}
\by{W. Hereman},
Symbolic software for Lie symmetry analysis, 
CRC Handbook of Lie group analysis of differential equations
\vol{3} Boca Raton, Florida: CRC Press (1996), 367-413, N.~H.~Ibragimov (ed).

\bibitem{Hussin}
\by{M. A. Ayari, V. Hussin}, GLie: A MAPLE program for Lie supersymmetries of Grassmann\/-\/valued differential equations, \jour{Comput.\ Phys.\ Commun.} \vol{100} (1997), 157-176.

\bibitem{Kersten}
\by{P. K. H. Gragert,  P. H. M. Kersten},
Differential geometric computations and computer algebra. Algorithms and software for symbolic analysis of nonlinear systems, \jour{Math.\ Comput.\ Modelling} \vol{25} (1997) no.8-9, 11--24.

\bibitem{Popowicz}
\by{Z. Popowicz}, SUSY2, \jour{Comput.\ Phys.\ Commun.} \vol{100} (1997), 277-296.

\bibitem{Krivonos}
\by{S. Krivonos, K. Thielemans}, A \textit{Mathematica} package for
computing $N=2$ superfield operator product expansions, \jour{Class.\
Quantum Grav.} \vol{13} (1996), 2899--2910.

\bibitem{WolfCrack}
\by{T. Wolf}, Applications of \textsc{Crack} in the classification of
integrable systems, \book{CRM Proc.\ Lecture Notes} \vol{37}
(2004), 283--300.\ \texttt{arXiv:nlin.SI/0301032}

\bibitem{SUSY}
\by{T. Wolf}, Supersymmetric evolutionary equations with higher order
symmetries (2003), \texttt{http://beowulf.ac.brocku.ca/\symbol{"7E}twolf/htdocs/susy/all.html}.

\bibitem{Dubna05}
\by{A. V. Kiselev, T. Wolf},
On weakly non\/-\/local, nilpotent, and super\/-\/recursion operators
for $N=1$ homogeneous super\/-\/equations,  Proc.\ Int.\ Workshop
`Supersymmetries and Quantum Symmetries--05,' JINR, Dubna
(July 27--31, 2005), 231--237.\ \texttt{arXiv:nlin.SI/0511056}

\bibitem{JKVin}
\by{I. S. Krasil'shchik, A. M. Vinogradov},
A method of calculating higher symmetries of nonlinear evolutionary equations, and nonlocal symmetries, \jour{Dokl.\ Akad.\ Nauk SSSR} \vol{253} (1980) no.6, 1289--1293.

\end{thebibliography}
\end{document}